\documentclass[showkeys,12pt, preprint,preprintnumbers,nofootinbib, groupedaddress,superscriptaddress,amsmath,amssymb]{revtex4}
\usepackage{graphicx}
\usepackage{dcolumn}
\usepackage{bm}
\usepackage{amssymb}
\usepackage{amsmath}
\usepackage{epsfig}    
\usepackage{color}
\usepackage{slashed}
\usepackage{hhline}
\usepackage{feynmp}

\def\be{\begin{equation}}
\def\ee{\end{equation}}
\newcommand{\bea}{\begin{eqnarray}}
\newcommand{\eea}{\end{eqnarray}}

\numberwithin{equation}{section}

\begin{document}

{\begin{flushright}{APCTP Pre2020 - 004}
\end{flushright}}

\title{Dirac dark matter in a radiative neutrino model} 
\author{Hiroshi Okada}
\email{hiroshi.okada@apctp.org}
\affiliation{Asia Pacific Center for Theoretical Physics (APCTP) - Headquarters San 31, Hyoja-dong,
Nam-gu, Pohang 790-784, Korea}
\affiliation{Department of Physics, Pohang University of Science and Technology, Pohang 37673, Republic of Korea}

\author{Yutaro Shoji}
\email{yshoji@kmi.nagoya-u.ac.jp}
\affiliation{Kobayashi-Maskawa Institute for the Origin of Particles and the Universe, Nagoya University, Nagoya, Aichi 464-8602, Japan}
\date{\today}

\begin{abstract}
We propose a simple radiative neutrino mass scenario with a Dirac dark
matter candidate, which is minimally realized by a $Z_3$ symmetry. We
introduce two Dirac neutrinos and two inert doublets. We demonstrate
that the model has a large allowed region that satisfies the constraints
from neutrino oscillation data, lepton flavor violations, direct/indirect
detection of dark matter and dark matter relic density. We also propose
an efficient parameterization of the neutrino Yukawa couplings, which
reproduces a desired active neutrino mass matrix.
\end{abstract}
\maketitle
\newpage

\section{Introduction}
It is very important to understand the nature of neutrinos such as their
oscillations, mass ordering, CP phases, Dirac or Majorana, and any
interactions related to neutrinos since they would have a clue to derive
new physics.  For example, the leptonic CP phases could explain the
Baryon Asymmetry of Universe (BAU) through
leptogenesis~\cite{leptogenesis}. It contrasts with the CP phase in the
quark sector, which is too small to explain the observed BAU.  Another
example is the neutrinoless double beta decay, which would tell us
whether neutrino is Dirac or Majorana, although the measurement of this
process is challenging.

Another hint of new physics is the existence of dark matter (DM), which
occupies $\sim27\%$ of the energy density of the
Universe~\cite{Aghanim:2018eyx}.  Although we have not discovered DM
yet, there are several suggestions from experiments.  For instance,
direct detection searches tell us an upper bound on the scattering cross
section of DM and nucleus. The most stringent bound is put by XENON1T
~\cite{Aprile:2018dbl} and is $\sigma_{SI}\lesssim 4.1\times
10^{-47}{\rm cm^2}$ at 30 GeV of DM mass for the spin independent
interaction, which is strong enough to exclude many models in this mass
range.  Meanwhile, indirect detection searches provide very attractive
signatures although there are rather big uncertainties from
astrophysical background.  Several interesting excesses in the
electron-positron flux of the cosmic rays are found by experiments such
as
AMS-02~\cite{Aguilar:2013pos,Accardo:2014pos,Aguilar:2014ele,Aguilar:2014all,Aguilar:2019ele,Aguilar:2019owu},
DAMPE~\cite{Ambrosi:2017all}, and
CALET~\cite{Adriani:2017all,Adriani:2018ktz}.  One of the
interpretations for them is to create electron-positron pair via DM
annihilation or decay. For example, let us consider a fermionic DM
candidate that annihilates into an electron-positron pair. Notice that the annihilation
channels to leptons naturally arise when the DM particle is involved in the diagram generating the
neutrino masses.  In this case, we need a few orders of magnitude larger
cross section than the canonical cross section of $\langle\sigma
v^2\rangle\simeq3\times10^{-26}{\rm cm^3/s}$ to explain these excesses.
One of the solutions is to consider a velocity dependent cross section
\cite{ArkaniHamed:2008qn,Bai:2014osa,Suematsu:2010gv}, which use
Sommerfeld~\cite{sf,Hisano:2004ds} or Breit-Wigner~\cite{Ibe:2008ye,
Guo:2009aj} to enhance the cross section at low temperature. Another
solution is to assume the existence of nearby DM sub-halos at 0.1-0.3
kpc away from the solar system
\cite{Yuan:2017ysv,Zu:2017dzm,Gu:2017gle,Duan:2017pkq,Fang:2017tvj,Fan:2017sor,Athron:2017drj,Chao:2017emq,Huang:2017egk,Tang:2017lfb,Ge:2017tkd}.
In either case, the Dirac feature of the DM is helpful since the annihilation
cross section is not velocity suppressed unlike the Majorana DM case.

In this paper, we propose a simple model to realize the neutrino mass
matrix and a Dirac DM in the framework of the radiative seesaw mechanism
at the one-loop level. We show an allowed region that satisfies the
constraints from neutrino oscillation data, lepton flavor violations
(LFVs), direct/indirect detection searches and correct relic density of
DM. Since the radiative seesaw scenario has been discussed in this
context for a long time, many authors have come up with various
models. Here, we list representative references at
one-loop~\cite{Ma:2006km,Hagedorn:2018spx},
two-loop~\cite{Kanemura:2011mw, Kajiyama:2013zla}, and
three-loop~\cite{Krauss:2002px, Aoki:2008av, Gustafsson:2012vj}.

This paper is organized as follows.  In Sec.~II, we explain our scenario
and formulate the neutrino sector and the LFVs.  In particular, we
provide an efficient parameterization for the neutrino Yukawa
couplings.  In Sect.~III, we discuss our DM candidate and give a
numerical analysis considering all the constraints mentioned above.
Finally, we summarize and conclude in Sec.~IV.

\section{Model setup}
\begin{widetext}
\begin{center} 
\begin{table}
\begin{footnotesize}
\begin{tabular}{|c||c|c|c||c|c|c|}\hline \hline 
 & \multicolumn{3}{c|}{Fermions} & \multicolumn{3}{c|} { Bosons} \\\hline
Fields  ~&~ $L_{L}$ ~ & ~$e_{R}$   ~&~ $N$ ~ &~ $H$ ~ &~ $H_1$ ~&~ $H_2$ ~
\\\hline 
 $SU(2)_L$  & $\bm{2}$  & $\bm{1}$  & $\bm{1}$   & $\bm{2}$  & $\bm{2}$   & $\bm{2}$ 
  \\\hline 
$U(1)_Y$  & $-\frac{1}{2}$ & $-1$ & $0$  & $\frac{1}{2}$ &  $\frac{1}{2}$  &  $\frac{1}{2}$ 
\\\hline
 $Z_{3}$ & $1$  & $1$ & $\omega$ & $1$  & $\omega$   & $\omega^2$ 
  \\\hline
\end{tabular}
\caption{The charge assignments of the relevant particles under
$SU(2)_L\times U(1)_Y\times Z_3 $, where $\omega\equiv e^{2\pi i/3}$.
We assume $N$ is Dirac and has two families and $H_1$ and $H_2$
do not develop VEVs.}  \label{tab:1}
 \end{footnotesize}
\end{table}
\end{center}
\end{widetext}

In this section, we review our scenario.  We introduce two families of
neutral Dirac fermions, $N$, to reproduce the neutrino data and two
inert doublets, $H_1=(\eta^+_1,\eta^0_1)^T$ and
$H_2=(\eta^+_2,\eta^0_2)^T$. The masses for $N_a$ is denoted as $M_a$
and those for $\eta^0_a$ and $\eta^+_a$ are denoted as $m_{\eta_a^0}$
and $m_{\eta_a^+}$ with $a\in\{1,2\}$, respectively. We denote the SM
Higgs boson as $H$ and its vacuum expectation value (VEV) as
$v_H/\sqrt2\simeq174$ GeV. In order to assure the Dirac nature of $N$
and the inert feature of $H_1$ and $H_2$, we impose a discrete Abelian
symmetry, $Z_3$. Under this rather simple framework, the neutrino mass
is induced at the one-loop level via the new particles and the lightest
particle among them can become a DM candidate. We assume the DM is the
lighter $N$, which we denote as $N_1$, and its stability is assured by
$Z_3$~\footnote{In ref.~\cite{Ma:2019coj}, this idea is proposed as one
of the possibilities.}.  We summarize the relevant field contents and
their charge assignments in Tab.~\ref{tab:1}.

The Lagrangian for the lepton sector is given by
\begin{align}
-{\cal L}_{L} &= y_{\ell_i} \bar L_{L_{i}}  H e_{R_{i}}
 + y_{N_{R_{ib}}} \bar L_{L_i} \tilde H_1 N_{R_b}
  + y_{N_{L_{aj}}} \bar N_{L_a} \tilde H_2 L^C_{L_j}
 +M_{a} \bar N_{L_a} N_{R_a} + {\rm c.c.},
\label{eq:lag-quark}
\end{align}
where $i,j\in\{1,2,3\}$, $a,b\in \{1,2\}$ and $\tilde H_{1,2}\equiv
i\sigma_2 H_{1,2}^*$ with $i\sigma_2$ being the complete anti-symmetric
matrix.  We assume that $y_\ell$ and $M$ are diagonal; we consider the
mass eigen-bases for the charged-leptons and the neutral Dirac fermions.

Meanwhile, the relevant part of the Higgs potential is given by
\begin{equation}
 V=\lambda_0(H^\dag H_1)(H^\dag H_2) + {\rm c.c.},
\end{equation}
where $\lambda_0$ plays an important rule in generating nonzero neutrino
masses by connecting $H_1$ and $H_2$ appropriately. We assume that
$\eta^0_1$ and $\eta^0_2$ are heavier than $1$ TeV and they are close to
the mass eigenstates; we treat $\lambda_0v^2$ as a small mixing
parameter assuming that $\lambda_0<4\pi$. In addition, we presume the neutral component
and the charged component of an inert doublet have almost degenerated
masses, {\it i.e.}  $m_{\eta_1^0}\simeq m_{\eta_1^+}\equiv m_{\eta_1}$
and $m_{\eta_2^0}\simeq m_{\eta_2^+}\equiv m_{\eta_2}$. Such larger
masses and small mixings are also preferable from the constraints on the
oblique parameters~\cite{Peskin:1990zt, Peskin:1991sw}, which become
important when the inert doublets have masses of several hundreds GeV
and large mass splittings between the charged and the neutral components
\cite{Grimus:2008nb}. We do not study the details of the oblique parameters since the corrections are expected to be small in our setup
and there are other tunable scalar couplings, {\it e.g.} $|H_1^\dagger
H|^2$, that are relevant for the oblique parameters but not related to
the constraints on neutrino masses, LFVs or dark matter properties.

\subsection{Active neutrino mass}
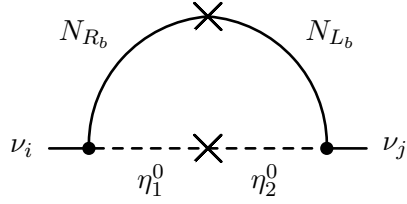
\begin{figure}[t]
\begin{fmffile}{mass}
\begin{fmfgraph*}(120,120)
\fmfleft{i}
\fmfright{o}
\fmftop{t}
\fmf{plain,tension=3}{i,v1}
\fmf{plain,tension=3}{v2,o}
\fmf{phantom,tension=0.01}{vp,t}
\fmflabel{$\nu_i$}{i}
\fmflabel{$\nu_j$}{o}
\fmf{plain,left=0.4,tension=0.001,label=$N_{R_b}$}{v1,vp}
\fmf{plain,left=0.4,tension=0.001,label=$N_{L_b}$}{vp,v2}
\fmf{dashes,label=$\eta^0_1$}{v1,vm}
\fmf{dashes,label=$\eta^0_2$}{vm,v2}
\fmfdot{v1,v2}
\fmfv{decor.shape=cross}{vm,vp}
\end{fmfgraph*}
\end{fmffile}
\caption{The one-loop diagram for the active neutrino mass. The diagram
with $i\leftrightarrow j$ contributes as well.}\label{fig_neutrino_mass}
\end{figure}

The dominant contribution to the active neutrino mass matrix comes from
the one-loop diagrams involving the Dirac neutral fermions and the
neutral inert Higgs bosons, which are shown in
Fig.~\ref{fig_neutrino_mass}.  At the leading order in $\lambda_0$, the
active neutrino mass matrix is given by~\cite{Okada:2015vwh}
\begin{align}
m_{\nu_{ij}}=y_{N_{R_{ib}}} R_{bb} y_{N_{L_{bj}}}+ y^T_{N_{L_{ib}}} R_{bb} y^T_{N_{R_{bj}}} ,
\end{align}
where the dimensionful parameter, $R$, is defined as
\begin{align}
R_{bb}&=\frac{\lambda_0^* v_H^2 M_b}{32\pi^2} F(M^2_b,m^2_{\eta^0_1},m^2_{\eta^0_2}),\\
F(a,b,c)&=-\frac{(b-c)a\ln a+(c-a)b\ln b+(a-b)c \ln c}{(a-b)(b-c)(c-a)}.
\end{align}

The neutrino mass matrix is diagonalized by a unitary matrix,
$U_\nu$, as
\begin{equation}
 U_\nu^T m_\nu U_\nu ={\rm diag}(m_1,m_2,m_3)\equiv D_\nu.
\end{equation}
We regard $U_\nu$ as the PMNS matrix, which we denote as $U_{\rm
PMNS}$~\cite{Maki:1962mu}, because the charged-lepton mass matrix is
assumed to be diagonal. We assume that $D_\nu$ does not have zero
eigenvalues in the following discussion\footnote{One may take a limit of a
vanishing eigenvalue in the following discussion, if one want to have a massless
neutrino.}.

Since there are a sufficient number of parameters, there exist
parameterizations that always reproduce the neutrino oscillation
data\footnote{The rank of the mass matrix for $N$ has nothing to do with
the rank of the active neutrino mass matrix since $y_{N_R}\neq
y_{N_L}^T$}. In the following, we show one of such parameterizations.

Let us first show the parameterization and explain the details later.
We define
\begin{align}
 x&=U_{\rm PMNS}^*\sqrt{D_\nu}(w^{(1)}+iw^{(2)}),\\
 y&=U_{\rm PMNS}^*\sqrt{D_\nu}w^{(3)},\\
 z&=U_{\rm PMNS}^*\frac{1}{\sqrt{D_\nu}}(w^{(1)}-iw^{(2)}),\label{eq_z}
\end{align}
where the elements of $\sqrt{D_\nu}$ are the square roots of the
elements of $D_\nu$, $1/\sqrt{D_\nu}$ is its inverse and $w^{(i)}$'s are
arbitrary real vectors with three components satisfying
\begin{equation}
 w^{(i)}\cdot w^{(j)}=\delta^{ij}.
\end{equation}

Then, we construct an orthonormal basis as
\begin{align}
 \hat x&=\frac{x}{|x|},\\
 \hat y&=\frac{y-\frac{x^\dagger y}{|y|^2}x}{\left|y-\frac{x^\dagger y}{|y|^2}x\right|},\\
 \hat z&=\frac{z}{|z|}.
\end{align}

Using them, the Yukawa couplings for the Dirac neutral fermions are
constructed as
\begin{align}
 y_{N_R}&=
\begin{pmatrix}
 \hat x&\hat y
\end{pmatrix}Y,\label{eq_ynr}\\
 y_{N_L}&=\frac{1}{2}R^{-1}Y^{-1}
\begin{pmatrix}
 \hat x^\dagger\\
 \hat y^\dagger
\end{pmatrix}
\left[U^*_{\rm PMNS}D_\nu U^\dagger_{\rm PMNS}+A\right]\label{eq_ynl},
\end{align}
where $Y$ is an arbitrary full rank $2\times2$ matrix and $A$ is an
anti-symmetric matrix satisfying\footnote{When $\hat z_2=0$, the square
parentheses in the right-hand sides of these equations need to be zero
simultaneously and $A_{12}$ and $A_{23}$ become free parameters. In the
numerical analysis, we choose $\hat z_2$ randomly. Thus, it does not
become strictly zero in practice.}
\begin{align}
 A_{12}&=-\frac{1}{\hat z^*_2}\left[A_{13}\hat z_3^*-(U^*_{\rm PMNS}D_\nu U^\dagger_{\rm PMNS}\hat z^*)_1\right],\label{eq_a12}\\
 A_{23}&=-\frac{1}{\hat z^*_2}\left[A_{13}\hat z_1^*+(U^*_{\rm PMNS}D_\nu U^\dagger_{\rm PMNS}\hat z^*)_3\right]\label{eq_a23},
\end{align}
where the subscripts indicate the index for the components.  Then, one
can easily check
\begin{equation}
 y_{N_{R}} R y_{N_{L}}+ y^T_{N_{L}} R y^T_{N_{R}}=U_{\rm PMNS}^*D_\nu U_{\rm PMNS}^\dagger,
\end{equation}
by using
\begin{equation}
 \hat z^\dagger\left[U^*_{\rm PMNS}D_\nu U^\dagger_{\rm PMNS}+A\right]=0.\label{eq_proj}
\end{equation}
In our numerical analysis, we take $A_{13}$, $Y$ and $w^{(i)}$'s as the
input parameters instead of $y_{N_R}$ and $y_{N_L}$.  We take into
account the perturbativity conditions, $|y_{N_{R_{ib}}}|<\sqrt{4\pi}$
and $|y_{N_{L_{bi}}}|<\sqrt{4\pi}$, and the cosmological constraint on
the sum of the neutrino masses, $\sum_i m_i\lesssim0.12$ eV
\cite{Aghanim:2018eyx, Vagnozzi:2017ovm}.

Let us explain the details of the above parameterization. First of all,
the most general $y_{N_R}Ry_{N_L}$ that reproduces the mass matrix can
be expressed as
\begin{equation}
 y_{N_R}Ry_{N_L}=\frac{1}{2}[U^*D_\nu U^\dagger +A],
\end{equation}
with $A$ being an arbitrary anti-symmetric matrix. Since $y_{N_R}$ and
$y_{N_L}^T$ are $3\times2$ matrices, $y_{N_R}Ry_{N_L}$ should have rank
2, {\it i.e.} there exists $\hat z$ so that $\hat z^\dagger
y_{N_R}Ry_{N_L}=0$, which gives the constraint of Eq.~\eqref{eq_proj}.
The necessary and sufficient conditions for the constraint are found to
be Eqs.~\eqref{eq_a12} and \eqref{eq_a23}. Notice that the special
parameterization of $z$ given in Eq.~\eqref{eq_z} is deduced from
$z^\dagger[U^*D_\nu U^\dagger+A]z^*=0$. Now, matrix $(\hat x~\hat y)$
becomes invertible on the plane perpendicular to $z$ and we can express
the Yukawa couplings as Eqs.~\eqref{eq_ynr} and \eqref{eq_ynl}.

\subsection{Lepton flavor violations (LFVs)} 
\begin{figure}[t]
\begin{fmffile}{lfv}
\begin{fmfgraph*}(120,120)
\fmfleft{i}
\fmfright{o}
\fmfright{0v1,op,ov2,ov3}
\fmf{plain,tension=4}{i,v1}
\fmf{plain,tension=4}{v2,o}
\fmf{photon,tension=0}{vm,op}
\fmflabel{$\ell_i$}{i}
\fmflabel{$\ell_j$}{o}
\fmflabel{$\gamma$}{op}
\fmf{plain,left,tension=0.4,label=$N_{R_a}(N_{L_a})$}{v1,v2}
\fmf{dashes}{v1,vm}
\fmf{dashes}{vm,v2}
\fmfv{label=$\eta^+_1(\eta^+_2)$,label.angle=90}{vm}
\fmfdot{v1,v2,vm}
\end{fmfgraph*}
\end{fmffile}
\caption{The one-loop diagram for LFV.}\label{fig_diag_lfv}
\end{figure}
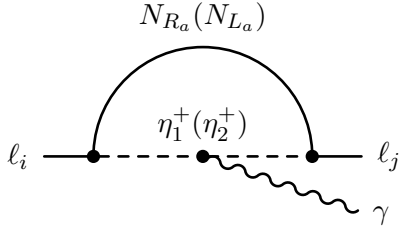

Due to the newly introduced couplings, $y_{N_R}$ and $y_{N_L}$, LFVs
arise at the one-loop level as shown in Fig.~\ref{fig_diag_lfv}.  The
branching ratios for $\ell_i\to\ell_j\gamma$ are calculated as
\begin{align}
&\frac{{\rm BR}(\ell_i\to \ell_j\gamma)}{{\rm BR}(\ell_i\to \ell_j\nu_j\nu_i)}=
 \frac{3\alpha_{\rm em}}{16\pi{\rm G_F^2}}
\left| y_{N_{R_{ja}}}y_{N_{R_{ia}}}^* G(M_a^2,m^2_{\eta^+_1}) 
+
y_{N_{L_{aj}}}y_{N_{L_{ai}}}^* G(M_a^2,m^2_{\eta^+_2})
 \right|^2,\label{eq_lfv}
\end{align}
where
\begin{equation}
 G(m_a^2,m_b^2)=\frac{2 m_a^6 +3 m_a^4 m_b^2 -6 m_a^2 m_b^4 + m_b^6+12 m_a^4 m_b^2 \ln(m_b/m_a)}{12(m_a^2-m_b^2)^4}.
\end{equation}
Here, ${\rm G_F}\approx 1.17\times10^{-5}$[GeV]$^{-2}$ is the Fermi
constant and $\alpha_{\rm em}\approx1/129$ is the fine structure
constant.  The current experimental upper bounds
are found as~\cite{TheMEG:2016wtm, Adam:2013mnn}
  \begin{align}
  {\rm BR}(\mu\rightarrow e\gamma) &\leq4.2\times10^{-13},\quad
{\rm BR}(\tau\rightarrow \mu\gamma)\leq4.4\times10^{-8},
\quad   {\rm BR}(\tau\rightarrow e\gamma) \leq3.3\times10^{-8}~.
 \label{expLFV}
 \end{align}

The same couplings also contribute to the muon anomalous magnetic moment.
Currently, there is a tention between the experimental value and the standard model prediction as \cite{Zyla:2020zbs,Hagiwara:2011af,
Keshavarzi:2018mgv, e989,
jpark,Davier:2010nc,Davier:2017zfy,Davier:2019can}
\begin{equation}
 \Delta a_\mu = a_\mu^{\rm exp}-a_\mu^{\rm SM}=261(63)(48)\times10^{-11},
\end{equation}
where $a_\mu^{\rm exp}$ and $a_\mu^{\rm SM}$ are the experimental value and the standard model prediction, respectively.
The first error comes from the experiment and the second comes from the theory.
Meanwhile, our model has additional contributions as
\begin{equation}
 \Delta a_\mu=-\frac{m_\mu^2}{8\pi^2}\left[|y_{N_{R_{\mu a}}}|^2 G(M_a^2,m^2_{\eta^+_1}) 
+
|y_{N_{L_{a\mu}}}|^2 G(M_a^2,m^2_{\eta^+_2})\right].
\end{equation}
Although the sign of the contribution is always negative \footnote{
$G(m_a^2,m_b^2)$ is always positive.}, its magnitude turned out to be smaller than about $10^{-12}$.
Thus, we do not discuss it furthermore in this paper.

\section{Dark matter}
In our model, the lightest particle among the Dirac fermions and the
inert doublets becomes a DM candidate.  In this paper, we assume the
lighter Dirac fermion, $\chi\equiv N_1$, is the DM and denote its mass
as $m_\chi\equiv M_1$. We also assume that the masses of the other $Z_3$
charged particles are not so close to the DM mass\footnote{If some of
the masses of the $Z_3$ charged particles become close to the DM mass,
one needs to include the coannihilation effects to get the correct relic
density. It typically requires degeneracy of $1\%-10\%$ in mass (see for
example
\cite{Boehm:1999bj,Ellis:1998kh,Baer:2005jq,Profumo:2004wk}). Since such
a region is rather fine-tuned in the parameter space, we do not consider
coannihilation in our analysis.}  and that there is no initial asymmetry
between the DM density and the anti-DM density.

\subsection{Relic density}
We consider the case where the DM relic density is explained by the
freeze-out mechanism. The relevant annihilation channels of the DM are
$\chi\chi\to\nu\nu,\ell\bar \ell$, which occur at the tree level.

In order to estimate the relic density, we solve the Boltzmann
equation, which is given by
\begin{equation}
 \frac{dn_\chi}{dt}+3Hn_\chi=-\langle\sigma|v_\chi|\rangle[n_\chi^2-(n_\chi^{{\rm eq}})^2],
\end{equation}
where $H$ is the Hubble parameter, $n_\chi$ is the number density of the
DM (not including the anti-DM), $n_\chi^{\rm eq}$ is the equilibrium number
density and $\langle\sigma|v_\chi|\rangle$ is the thermally averaged
cross section, which is given in Appendix \ref{apx_th_xs}.

Then, we compare it with the observed relic density at $2\sigma$, which
is given by~\cite{Aghanim:2018eyx}
\begin{align}
\Omega_\chi h^2=0.1193\pm0.0018.
\end{align}

\subsection{Direct detection}
Next, we discuss the direct detection searches for the DM. We consider
only the spin independent (SI) processes since the spin-dependent ones
do not give a stringent constraint.  The DM-Higgs coupling is induced at
one-loop level through $y_{N_R}$ and $y_{N_L}$. In order to estimate the
scattering cross section, we write down the relevant interactions as
\begin{align}
-{\cal L}=\kappa^0_1 v_H h|\eta^0_1|^2+\kappa^+_1 v_H h|\eta^+_1|^2+
\kappa^0_2 v_H h|\eta^0_2|^2+\kappa^+_2 v_H h|\eta^+_2|^2,
\end{align}
where $h$ is the SM Higgs boson.
The effective couplings of the DM to the quarks are calculated as
\begin{align}
-{\cal L}_{\rm eff}&=C_{\rm SI} m_q \left(\bar\chi\frac{-i\partial_\mu\gamma^\mu}{m_\chi}\chi\right)(\bar qq),\\
C_{\rm SI}&=\frac{1}{32\pi^2}\frac{1}{m_h^2 m_\chi}
\left[|y_{N_{R_{a1}}}|^2\left(\kappa^0_1 I(m^2_{\eta^0_1},m^2_\chi)
+\kappa^+_1 I(m^2_{\eta^+_1},m^2_\chi)\right)
\right.\nonumber\\
&\hspace{17ex}\left.+|y_{N_{L_{1a}}}|^2\left(\kappa^0_2 I(m^2_{\eta^0_2},m^2_\chi)
+\kappa^+_2 I(m^2_{\eta^+_2},m^2_\chi)\right)
\right],\\
I(a,b)&=1-\frac{a-b}{b}\ln\frac{a}{a-b},
\end{align}
where $a>b$.
Then, the SI cross section is given by
\begin{align}
\sigma_{\rm SI}^p=\frac{m_p^2}{\pi}\left(\frac{m_\chi m_p}{m_\chi+ m_p}\right)^2|C_{\rm SI}|^2 f_p^2,
\end{align}
where $f_p=0.326$~\cite{Mambrini:2011ik}.  Since the newly introduced couplings,
$\kappa_1^0$, $\kappa_1^+$, $\kappa_2^0$ and $\kappa_2^+$, are not
related to the neutrino masses, the LFVs or the DM relic density, we
cannot get a rigid prediction for the cross section. Instead, we
consider the maximum value of the cross section with
\begin{equation}
 |\kappa_1^0|,|\kappa_1^+|,|\kappa_2^0|,|\kappa_2^+|<\kappa_{\max}.
\end{equation}
The maximum value is evaluated as
\begin{equation}
 \sigma_{\rm SI}^p<\sigma_{\rm SI}^{p,\max}=\kappa_{\max}^2\sigma_{\rm SI}^p|_{\kappa_1^0\to1,\kappa_1^+\to1,\kappa_2^0\to1,\kappa_2^+\to1}.
\end{equation}
To discuss the maximum detectability of this model, we compare
$\sigma_{\rm SI}^{p,\max}$ with the current experimental constraints
such as that of XENON1T~\cite{Aprile:2018dbl}, which provides the most stringent constraint, as
well as the expected sensitivity of LZ..

\begin{figure}[t]
\begin{fmffile}{direct}
\begin{fmfgraph*}(120,60)
\fmfleft{i1,i2}
\fmfright{o1,o2}
\fmflabel{$q$}{i1}
\fmflabel{$\chi$}{i2}
\fmflabel{$q$}{o1}
\fmflabel{$\chi$}{o2}
\fmf{fermion}{i1,v1,o1}
\fmf{dashes,tension=0.6,label=$\eta_a^+(\eta_a^0)$,l.s=right}{v3,v2}
\fmf{dashes,tension=0.6}{v2,v4}
\fmf{fermion}{i2,v3}
\fmf{fermion,label=$\ell_i(\nu_i)$,l.s=left}{v3,v4}
\fmf{fermion}{v4,o2}
\fmf{dashes,label=$h$}{v1,v2}
\end{fmfgraph*}
\end{fmffile}\hspace{9ex}
\begin{fmffile}{indirect}
\begin{fmfgraph*}(120,60)
\fmfleft{i1,i2}
\fmfright{o1,o2}
\fmflabel{$\chi$}{i1}
\fmflabel{$\chi$}{i2}
\fmflabel{$\ell_i$}{o1}
\fmflabel{$\ell_j$}{o2}
\fmf{fermion}{i1,v1,o1}
\fmf{fermion}{i2,v3,o2}
\fmf{dashes,label=$\eta^+_a$}{v1,v3}
\end{fmfgraph*}
\end{fmffile}
\caption{The diagrams for direct (left) and indirect (right) detection.}\label{fig_direct/indirect}
\end{figure}
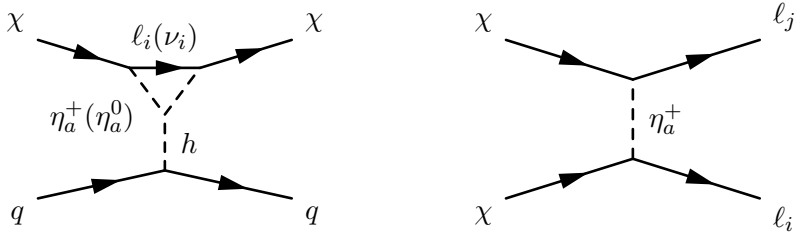

\subsection{Indirect detection}
The signals of the DM annihilating into leptons have been searched by
observing cosmic rays, whose flux has been recorded by Fermi-LAT, AMS-02,
DAMPE and CALET. In particular, an interesting excess in the
electron-positron flux has been observed and attracted attention. In
addition, the annihilation into charged leptons cause diffuse emission
of gamma rays, which have been measured by Planck, Fermi-LAT, and
H.E.S.S.

For the DM mass larger than a few hundred GeV, which is the relevant
mass region in our setup, H.E.S.S. gives the most stringent constraints
\cite{Abdallah:2016ygi}.  In our analysis, we assume that one of
$e^+e^-,\mu^+\mu^-$ and $\tau^+\tau^-$ dominates the annihilation
channels, which is a good approximation unless the largest elements of
$y_{N_L}$ and $y_{N_R}$ are accidentally close to each other\footnote{
    When some of the elements are close to each other, we need a more sophisticated analysis due to multiple decay channels.
    Since it happens only in finetuned regions, we ignore this effect in our analysis.
}. We
evaluate the annihilation cross sections using the formulas in Appendix
\ref{apx_th_xs} with $T\sim0$.

\subsection{Numerical analysis}
Here, we perform a numerical analysis to find allowed regions, where the
relic density of DM, the neutrino oscillation data and the LFV
constraints are satisfied. First of all, we fix the ranges for input
parameters as
\begin{align}
&|\lambda_0|<4\pi,\quad 0\le \alpha_{2,3}< 360^\circ,\\
&m_0<m_0^{\max},\quad M_1<M_2<10\ {\rm TeV},\\
&M_1< m_{\eta_{1,2}},\quad 1\ {\rm TeV}< m_{\eta_{1,2}}< 10\ {\rm TeV},\\
&|Y_{ab}|<\sqrt{4\pi},\quad |A_{13}|<10\ {\rm eV},
\end{align}
where $m_0$ is the (non-vanishing) lightest neutrino mass and
$\alpha_{2,3}$ are the Majorana phases. Here, $m_0^{\max}$ is the
maximum value of the lightest neutrino mass to satisfy the cosmological
constraint on the sum of the neutrino mass, $\sum_i m_i\lesssim0.12~{\rm
eV}$. We have $m_0^{\max}=0.0301~{\rm eV}$ for the normal hierarchy and
$m_0^{\max}=0.0157~{\rm eV}$ for the inverted hierarchy.

For the neutrino oscillation parameters, we use the central values given by \cite{Esteban:2018azc}
\begin{align}
& \theta_{12}=33.82^\circ,\quad \theta_{23}=49.7^\circ,\quad \theta_{13}=8.61^\circ,\quad\delta_{\rm CP}=217^\circ,\\
& \Delta m^2_{21}=7.39\times 10^{-5}~{\rm eV^2 },\quad
\ \Delta m^2_{31}=2.525\times 10^{-3}~{\rm eV^2},
\end{align}
for the normal hierarchy and
\begin{align}
& \theta_{12}=33.82^\circ,\quad \theta_{23}=49.7^\circ,\quad \theta_{13}=8.65^\circ,\quad\delta_{\rm CP}=280^\circ,\\
& \Delta m^2_{21}=7.39\times 10^{-5}~{\rm eV^2 },\quad
\ \Delta m^2_{32}=-2.512\times 10^{-3}~{\rm eV^2},
\end{align}
for the inverted hierarchy. Here, $\Delta m_{ij}^2=m_{\nu_i}^2-m_{\nu_j}^2$.

\begin{figure}[!hptb]
\begin{center}
\includegraphics[width=70mm]{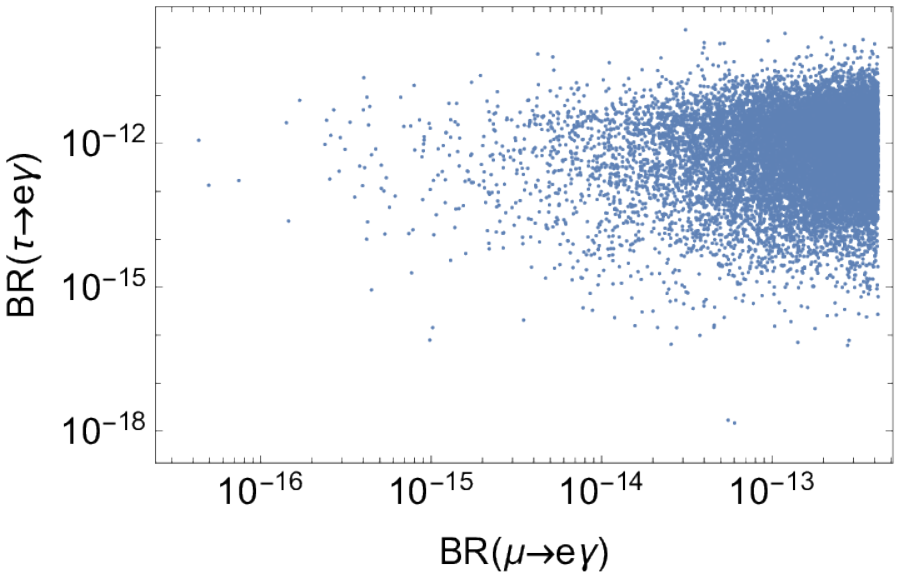} \qquad
\includegraphics[width=70mm]{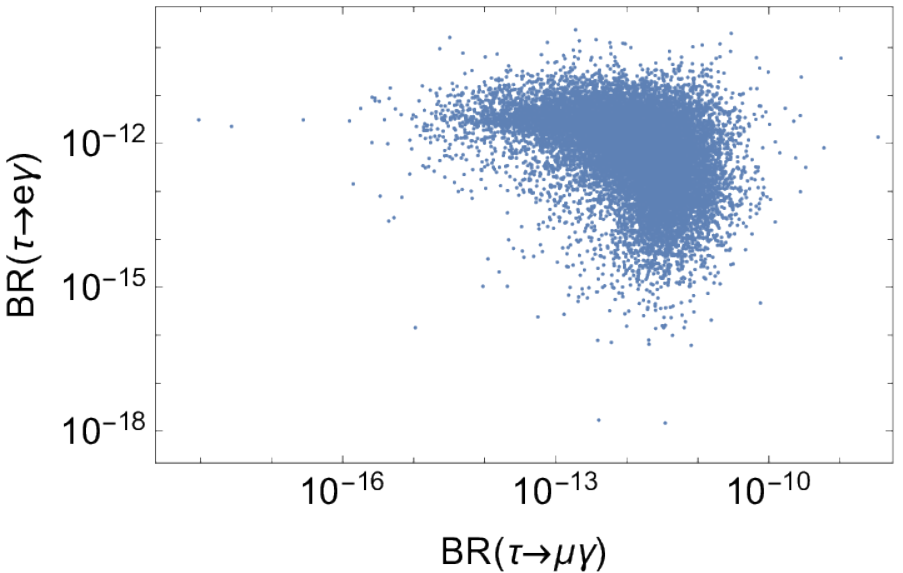}\caption{The branching ratios of
LFV processes for the normal hierarchy. }  \label{fig_lfv}
\end{center}\end{figure}

In Fig.~\ref{fig_lfv}, we show the scatter plots on the plane of
LFVs. All the points satisfy the constraints on the LFVs and the DM
relic density. Notice that the neutrino mixing angles and mass gaps are
fixed to the central values.  We can see that the constraint on $\mu\to
e\gamma$ is the most stringent and the other LFVs are about two or more
orders of magnitude smaller than the current constraints.

\begin{figure}[!hptb]
\begin{center}
\includegraphics[width=70mm]{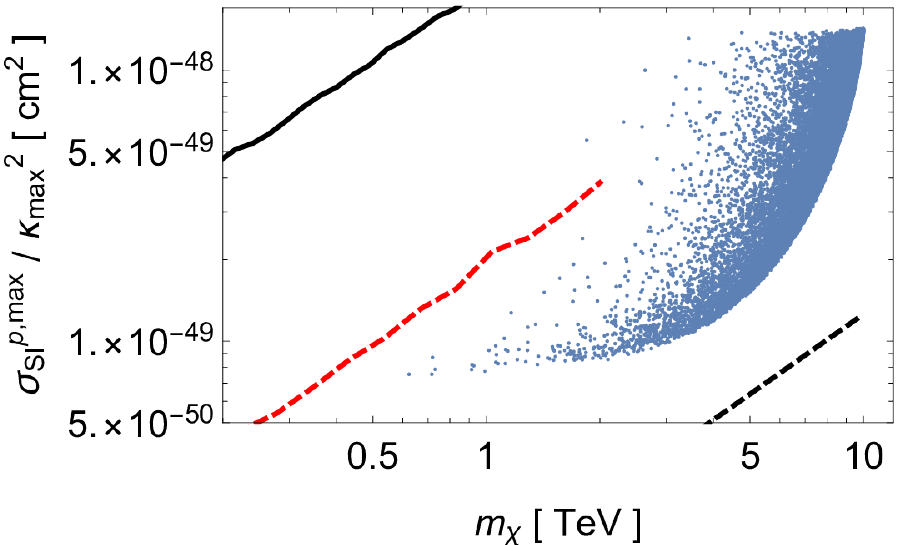} \qquad
\includegraphics[width=70mm]{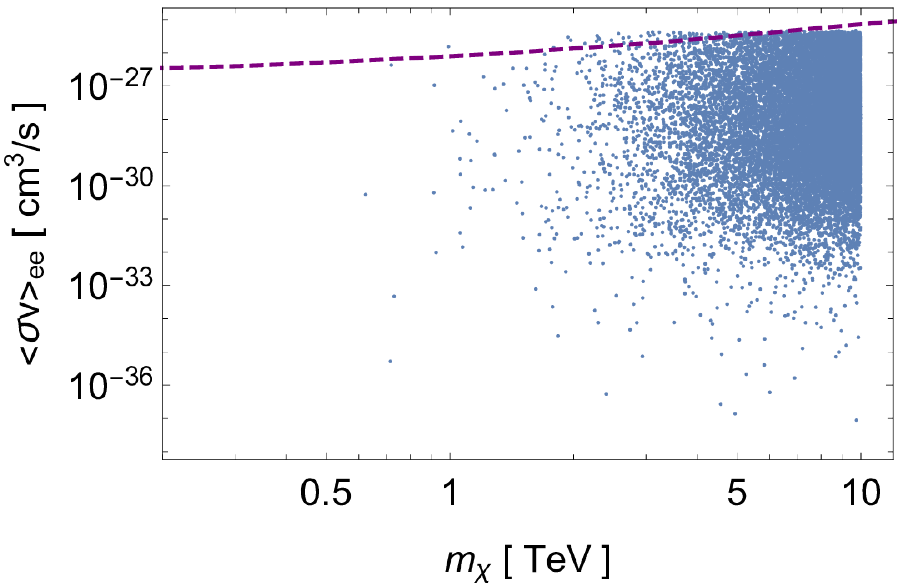}
\includegraphics[width=70mm]{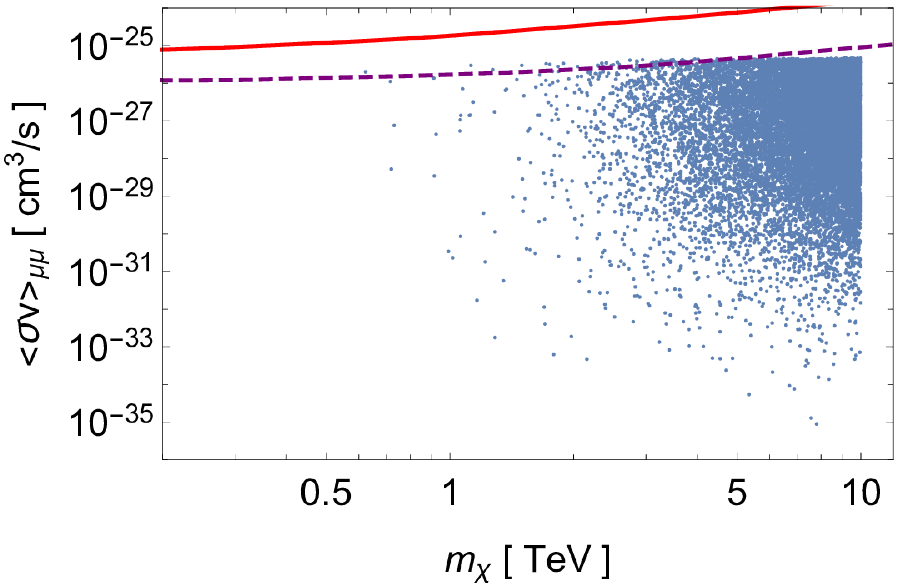} \qquad
\includegraphics[width=70mm]{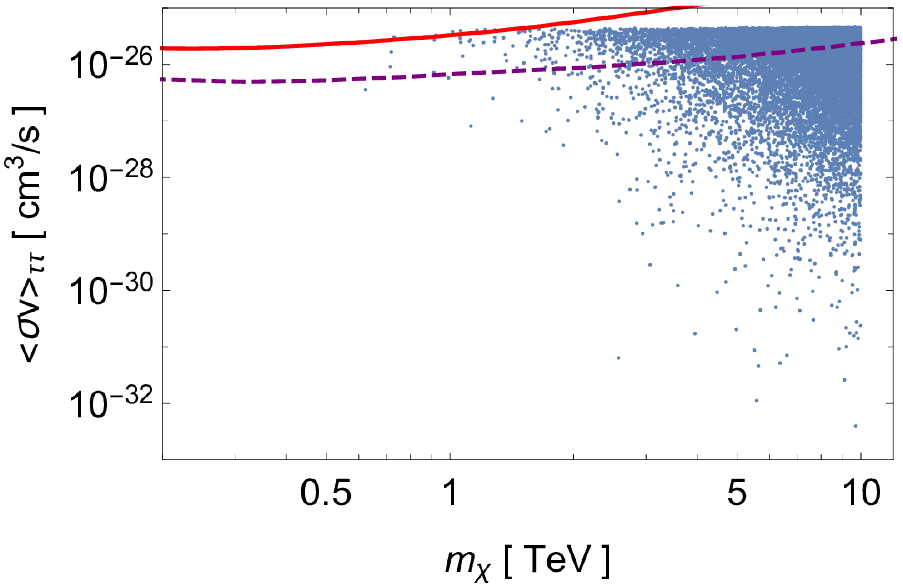}\caption{Constraints from the
direct detection and the indirect detection for the normal
hierarchy. The upper left panel shows the cross section for the direct
detection. The black solid (dashed) line corresponds to the coherent
neutrino scattering with $\kappa_{\max}=1(4\pi)$, respectively. The
expected sensitivity of LZ with $\kappa_{\max}=4\pi$ is shown with red
dashed line. The upper right panel shows the thermal average of the
cross section into $e^+e^-$ and the bottom left and right panels show
those into $\mu^+\mu^-$ and $\tau^+\tau^-$, respectively. The red solid line
corresponds to the H.E.S.S. constraint. The purple dashed lines show the sensitivity of CTA observing the Galactic
halo for 500h.} \label{fig_search}
\end{center}\end{figure}

In the top left panel of Fig.~\ref{fig_search}, we show the constraints
from direct searches for the normal hierarchy. The black solid
(dashed) line shows the coherent neutrino scattering with
$\kappa_{\max}=1(4\pi)$, respectively. The red dashed line shows the
expected sensitivity of LZ \cite{Mount:2017qzi} \footnote{
    The red dashed line ends at $2$ TeV due to the range of the original figure in \cite{Mount:2017qzi}.
}
with
$\kappa_{\max}=4\pi$. Since the current limit from XENON1T is much above the scattered points even with $\kappa_{\max}=4\pi$,
we do not plot the corresponding lines.
As we can see, the cross section is too small for
$\kappa_{\max}=1$ to be detected at the direct detection
experiments. However, with $\kappa_{\max}=4\pi$, which is almost the
maximum value considering the perturbativity, the points are still above
the coherent neutrino scattering and the DM signals may be observed at
the future experiments
\footnote{We need $\kappa_{\max}\gtrsim2.5$ to make some points above the coherent neutrino scattering.}.

In the top right and bottom panels, we show the thermal average of the
cross sections into $e^+e^-$, $\mu^+\mu^-$ and $\tau^+\tau^-$ for the
indirect detection. We show the constraint from H.E.S.S. \cite{Abdallah:2016ygi} with the red
line \footnote{The limit on the $e^+e^-$ channel is not available in \cite{Abdallah:2016ygi}.}. As we can see, the maximum value of the cross section is around
the nominal cross section of $\langle\sigma
v\rangle=3\times10^{-26}~{\rm cm^3/s}$ as is expected. When
$\tau^+\tau^-$ channel dominates the cross section and the DM mass is
smaller than $1.5$ TeV, the points are excluded by H.E.S.S. We also show
the prospect of CTA \cite{Lefranc:2015pza,Morselli:2017ree} observing the Galactic halo
for 500h with the purple dashed lines.

In Figs~\ref{fig_lfv} and \ref{fig_search}, we show only the normal
hierarchy since we get similar plots for the inverted
hierarchy. This is because we have free parameters, $A_{13}$, $Y$, $\hat
x$ and $\hat y$, in Eq.~\eqref{eq_ynl}, which hide the difference coming
from the mass hierarchy. In both hierarchies, we find a large number of
points that satisfy all the constraints we have discussed.

\section{Conclusions and discussions}
We have studied a simple neutrino mass scenario at the one-loop level,
which also contains a Dirac DM candidate. The model is minimally
realized by two Dirac neutrinos and two inert doublets with a $Z_3$
symmetry. In formulating the neutrino sector, we have developed an
efficient method to parameterize the neutrino Yukawa couplings.  We have
shown an allowed region that satisfies the constraints from neutrino
oscillation data, lepton flavor violations, direct/indirect detection of
dark matter and dark matter relic density. We have found that the
allowed region covers almost whole the range of DM mass we have assumed,
$0.2-10$ TeV. With an enhancement mechanism of the cross section or the
existence of DM sub-halos, it may also explain the excesses in the
electron-positron flux as discussed in
\cite{ArkaniHamed:2008qn,Bai:2014osa,Tang:2017lfb,Ge:2017tkd}.  The
simplest realization is to make use of the Breit-Wigner enhancement,
introducing a singlet boson that has a vanishing VEV and couples to DM
pairs.  Then, the DM annihilates into lepton-pairs via one-loop diagrams induced by the mixing between the singlet
boson and the SM Higgs boson~\cite{Suematsu:2010gv}.
Considering that such models typically require flavor specific final
state, {\it e.g.} into an electron positron pair, it could lead to a
verifiable scenario by introducing flavor dependent symmetries such as a gauged electron minus muon $U(1)$ symmetry~\cite{Motz:2020ddk}.
Although we did not give a detailed analysis on these excesses, it would
be worthwhile to mention this issue since the Dirac nature of DM leads to a $s$-wave dominant contribution in the annihilation cross section. It might
be a hint of the large annihilation cross section, which is needed for the explanation of the positron-electron spectrum reported by AMS-02, DAMPE and CALET.

\begin{acknowledgements}
This research was supported by an appointment to the JRG Program at the
APCTP through the Science and Technology Promotion Fund and Lottery Fund
of the Korean Government. This was also supported by the Korean Local
Governments - Gyeongsangbuk-do Province and Pohang City (H.O.).
H.O.~is sincerely grateful for the KIAS member. Y.S.~is supported by
Grant-in-Aid for Scientific research from the Ministry of Education,
Science, Sports, and Culture (MEXT), Japan, No.~16H06492.
\end{acknowledgements}

\appendix
\section{Dark matter cross section}\label{apx_th_xs}
In this appendix, we give the thermally averaged cross sections for the
DM, which is used for the calculation of the thermal relic density.

For the $\chi\bar\chi\to2\nu$ processes, we have
\begin{align}
 \langle \sigma_{\chi\bar{\chi}\to 2\nu}|v_\chi|\rangle&=\sum_{a,b}\left(\frac{1}{n_{\chi}^{\rm eq}}\right)^2\frac{T}{512\pi^5}\int_{4m_\chi^2}^\infty ds~K_1\left(\frac{\sqrt{s}}{T}\right)\sqrt{s-4m_\chi^2}\nonumber\\
 &\hspace{3ex}\times\left\{|y_{N_{R_{a1}}}y_{N_{R_{b1}}}^*|^2W_1(m_{\eta_1^0}^2,s)\right.\nonumber\\
 &\hspace{8ex}+|y_{N_{L_{1a}}}^*y_{N_{L_{1b}}}|^2W_1(m_{\eta_2^0}^2,s)\nonumber\\
 &\left.\hspace{8ex}-\left(y_{N_{R_{a1}}}y_{N_{R_{b1}}}^*y_{N_{L_{1b}}}y_{N_{L_{1a}}}^*+c.c.\right)\frac{m_\chi^2}{s}W_2(m_{\eta_1^0}^2,m_{\eta_2^0}^2,s)\right\}.
\end{align}
For the $\chi\bar\chi\to \ell\bar \ell$ processes, we have
\begin{align}
 \langle \sigma_{\chi\bar{\chi}\to \ell\bar \ell}|v_\chi|\rangle&=\sum_{a,b}\left(\frac{1}{n_{\chi}^{\rm eq}}\right)^2\frac{T}{512\pi^5}\int_{4m_\chi^2}^\infty ds~K_1\left(\frac{\sqrt{s}}{T}\right)\sqrt{s-4m_\chi^2}\nonumber\\
 &\hspace{3ex}\times\left\{|y_{N_{R_{a1}}}y_{N_{R_{b1}}}^*|^2W_1(m_{\eta_1^+}^2,s)\right.\nonumber\\
 &\hspace{8ex}+|y_{N_{L_{1b}}}^*y_{N_{L_{1a}}}|^2W_1(m_{\eta_2^+}^2,s)\nonumber\\
 &\left.\hspace{8ex}-\left(y_{N_{R_{a1}}}y_{N_{R_{b1}}}^*y_{N_{L_{1b}}}y_{N_{L_{1a}}}^*+c.c.\right)\frac{m_\chi^2}{s}W_2(m_{\eta_1^+}^2,m_{\eta_2^+}^2,s)\right\}.
\end{align}
Here, $K_1(x)$ is the modified Bessel function of the second kind and
\begin{align}
 W_1(m_1^2,s)&=1+\frac{(a-1)^2}{a^2-\beta^2}-\frac{a-1}{\beta}\ln\frac{a+\beta}{a-\beta},\\
 W_2(m_1^2,m_2^2,s)&=\frac{2}{\beta(a+b)}\ln\frac{(a+\beta)(b+\beta)}{(a-\beta)(b-\beta)},
\end{align}
with
\begin{align}
 a&=\frac{2(m_1^2-m_\chi^2)}{s},\\
 b&=\frac{2(m_2^2-m_\chi^2)}{s},\\
 \beta&=\sqrt{1-\frac{4m_\chi^2}{s}},
\end{align}
for $a>1$ and $b>1$.

\end{document}